\documentclass[10pt,final,journal,letterpaper,oneside,twocolumn,]{IEEEtran}

\usepackage{cite}
\usepackage[pdftex]{graphicx}
\usepackage[cmex10]{amsmath}
\usepackage{array}
\ifCLASSOPTIONcompsoc
  \usepackage[caption=false,font=normalsize,labelfont=sf,textfont=sf]{subfig}
\else
  \usepackage[caption=false,font=footnotesize]{subfig}
\fi
\usepackage{dblfloatfix}
\usepackage{url}
\begin{document}

\title{Beam Tests of Beampipe Coatings for Electron Cloud Mitigation in Fermilab Main Injector}

\author{Michael~Backfish,
		Jeffrey~Eldred,~\IEEEmembership{Member,~IEEE,}
		Cheng-Yang Tan,
		and Robert~Zwaska,~\IEEEmembership{Senior Member,~IEEE,}%
\thanks{Operated by Fermi Research Alliance, LLC under Contract No. DE-AC02-07CH11359 with the United States Department of Energy.}
\thanks{The authors are with the Fermi National Accelerator Facility, Batavia, IL, 60510 USA (email: backfish@fnal.gov; jseldred@fnal.gov; cytan@fnal.gov; zwaska@fnal.gov).}%
\thanks{J. Eldred is also with the Department of Physics, Indiana University, Bloomington, IN, 47405 USA.}%
\thanks{Manuscript received April 23, 2015.}}

\maketitle

\begin{abstract}
Electron cloud beam instabilities are an important consideration in virtually all high-energy particle accelerators and could pose a formidable challenge to forthcoming high-intensity accelerator upgrades. Dedicated tests have shown beampipe coatings dramatically reduce the density of electron cloud in particle accelerators. In this work, we evaluate the performance of titanium nitride, amorphous carbon, and diamond-like carbon as beampipe coatings for the mitigation of electron cloud in the Fermilab Main Injector.  Altogether our tests represent 2700 ampere-hours of proton operation spanning five years. Three electron cloud detectors, retarding field analyzers, are installed in a straight section and allow a direct comparison between the electron flux in the coated and uncoated stainless steel beampipe. We characterize the electron flux as a function of intensity up to a maximum of 50 trillion protons per cycle. Each beampipe material conditions in response to electron bombardment from the electron cloud and we track the changes in these materials as a function of time and the number of absorbed electrons. Contamination from an unexpected vacuum leak revealed a potential vulnerability in the amorphous carbon beampipe coating. We measure the energy spectrum of electrons incident on the stainless steel, titanium nitride and amorphous carbon beampipes. We find the electron cloud signal is highly sensitive to stray magnetic fields and bunch-length over the Main Injector ramp cycle. We conduct a complete survey of the stray magnetic fields at the test station and compare the electron cloud signal to that in a field-free region.

\end{abstract}


\section{Introduction}

\IEEEPARstart{E}{lectron} cloud instabilities have been observed in proton beams at the Super Proton Synchrotron (SPS)~\cite{arduini}, Large Hadron Collider (LHC)~\cite{iadarola2014}, Relativistic Heavy Ion Collider (RHIC)~\cite{fischer2008}, Proton Storage Ring (PSR)~\cite{macek2001}, and Spallation Neutron Source (SNS) Accumulator Ring~\cite{blask2003}. Electron cloud was first observed in the Fermilab Main Injector in 2006~\cite{Xiaolong2007} using an retarding field analyzer (RFA) designed by Richard Rosenberg~\cite{Rosenberg2000}. Electron cloud might also be responsible for a recent instability observed in the Fermilab Recycler~\cite{eldredHB,adamson}.

The formation of electron cloud buildup in a beampipe is initiated by either of two main mechanisms in proton synchrotons~\cite{zwaska2011}. Firstly, high energy particle beams can ionize the residual gases within the vacuum chamber and free the electrons. Additionally, loss particles from the beam can scatter electrons as they are absorbed by the beampipe. The non-relativistic electrons freed by these mechanisms, can receive a net transverse acceleration via the Lorentz force of a passing beam bunch. These accelerated electrons collide with the beampipe and scatter a larger number of electrons via secondary electron emission. This new group of electrons can repeat the process with more electrons generated at every bunch passing. This avalanche process can build up an electron cloud of sufficient density to cause instabilities in the beam.

The density of the electron cloud depends critically on beam intensity and secondary electron yield (SEY) of the inner surface of the beampipe ~\cite{zwaska2011,lebrun2010,furman2011}. A reduction of the secondary electron yield of the inner beampipe surface would allow the beam intensity to be increased while maintaining or reducing the density of the electron cloud.

One promising way to reduce electron cloud formation is to coat the inside of the beampipe with low-SEY materials. A test in the KEKB Low Energy Ring (LER) has found titanium nitride (TiN) and diamond-like carbon (DLC) to have 1/3 and 1/5 of the electron cloud current as copper beampipe~\cite{kato2010} when exposed to a positron beam. The performance of beampipe coatings for mitigating electron cloud in positron beams has also been tested at the Cornell Electron Strorage (CESR)~\cite{calvey,CalveyRFA} and at the Stanford Linear Accelerator (SLAC) Low Energy Ring (LER)~\cite{pivi}. A test at the CERN Super Proton Synchrotron (SPS) has found almost complete suppression of electron cloud with amorphous carbon ({a-C}) coatings~\cite{vallgren2011} and found DLC to have 1/3 of the electron cloud current as stainless steel~\cite{vallgrenIPAC}. Proposals have been made to apply a {a-C} coating to the beampipe of CERN SPS by 2019, after commissioning Linac4 ~\cite{gilardoni2012,costa2014}.

In this paper, we test TiN, {a-C}, and DLC coatings in the Fermilab Main Injector and compare the performance of each coating to the performance of the uncoated stainless steel. We do not find complete suppression of electron cloud with the {a-C} beampipe coating and we also found results suggesting that the {a-C} coating may be vulnerable to contamination when exposed to air. In contrast, we found the performance of the DLC coatings to be superior to that found in other accelerator tests.

The current parameters of the Main Injectors are given in Table~\ref{Params}. Fermilab is also implementing an accelerator upgrade, known as the Proton Improvement Plan II (PIP-II)~\cite{pipII}, to increase the Main Injector beam intensity by 50\% (to {$75 \times 10^{12}$} protons) and be completed in 2023. The PIP-II upgrade is also designed to be compatible with a proposed future upgrade to double the Main Injector beam intensity (to {$150 \times 10^{12}$} protons) by replacing the Fermilab Booster~\cite{nagaitsev}.

\begin{table}[!t]
\caption{Main Injector parameters and beampipe diameters}
\label{Params}
\centering
\begin{tabular}{| l | l |}
\hline
Energy                                  & $8$--$120$ GeV \\
Circumference                           & $3319.4$ m \\
RF frequency                            & $52.8$--$53.1$ MHz \\
Beam Intensity                            & $20$--$50 \times 10^{12}$ protons \\ 
%Bunch Intensity                         & $12\times 10^{10}$ (slip-stacked) \\
Bunch Spacing                           & $18.9$ ns \\
Bunch Length                            & $1$--$10$ ns $@$ $95$\%\\
Beam Admittance                          & $40\pi$~mm$\cdot$mrad \\
Beam Emittance                          & $15\pi$~mm$\cdot$mrad \\
\hline
Beampipe Inner Diameter\hspace{.2cm}    &         \\
\hspace{0.5cm} steel, TiN and DLC         & $149.2$ mm \\ 
\hspace{0.5cm} {a-C}                         & $155.0$ mm \\  
\hline
\end{tabular}
\end{table}

\section{Experimental Setup}

The beampipe test station at MI-52~\cite{rookieMI} was installed in a drift region of the Main Injector in 2009. One section of stainless-steel beampipe has no coating and the other section is coated with a test material. The drift region extends 5 meters downstream of the coated beampipe and 1 meter upstream of the uncoated beampipe (where a quadrupole magnet is located). The first run evaluated performance of the TiN beampipe coating from September 11, 2009 to July 4, 2010. The second run evaluated performance of the {a-C} beampipe coating from August 23, 2010 to January 10, 2011. The third run began on September 12, 2013 and the third run data presented in this paper ends on September 10, 2014. All data currently collected for the third run has been at lower intensity ({$\sim\! 25\times 10^{12}$ }protons) and therefore, we expect to continue the third run for another year as high-intensity data becomes available. During the first and second runs, slip-stacking took place in the Fermilab Main Injector~\cite{slipStack1,brown}. Currently slip-stacking is being commissioned in the Fermilab Recycler for higher power operation and the per-pulse intensity is limited until commissioning is complete~\cite{adamson}.

\begin{figure}[!t]
   \centering
     \includegraphics[scale=0.4]{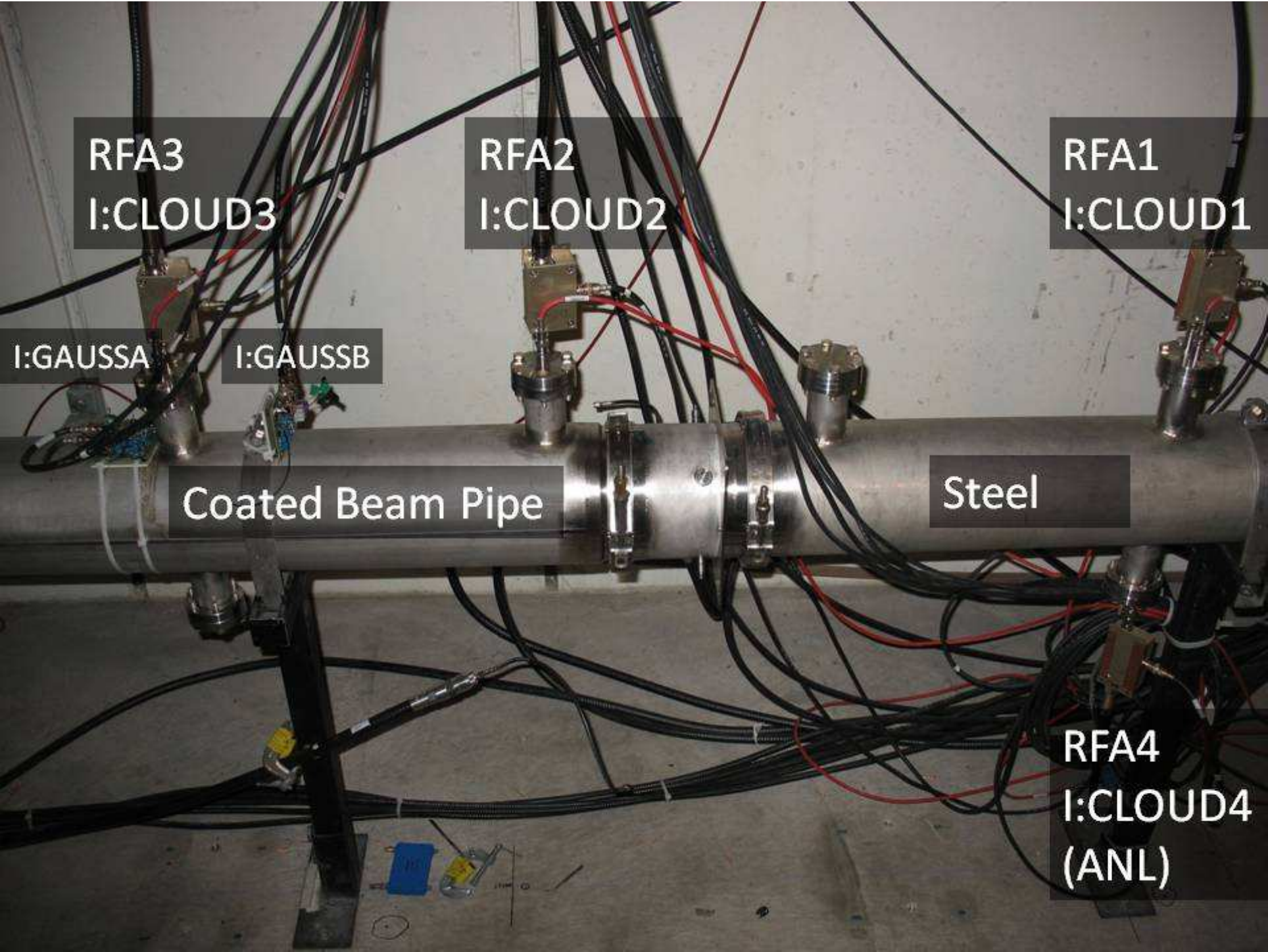}       
     \caption{The electron cloud measurement setup in the Main Injector at MI-52. The setup primarily consists of four RFAs and two beampipe sections. The beampipe is $6{\hbox{\tt "}}$ in diameter and the coated and uncoated sections are each $\sim$1 meter long. The setup is located in a drift region to avoid electron confinement from magnets. Stray magnetic fields are analyzed in section~\ref{MagSec}.}
     \label{MI52}
\end{figure}

Four electron cloud detectors, called retarding field analyzers (RFAs), are installed at the MI-52 to compare the electron flux at three locations as indicated in Fig.~\ref{MI52}. RFA1 is located in the uncoated stainless steel section. RFA2 is located in the coated beampipe section, 5 cm downstream from the boundary between the uncoated and coated beampipe surfaces. RFA3 is located in center of the coated beampipe section, 46 cm in either direction from the boundary between the uncoated and coated beampipe surfaces. RFA1 measures the electron cloud of the control group, RFA3 measures the electron cloud of the treatment group, and RFA2 measures the electron cloud of the transition region.

\begin{figure}[!t]
 \centering
  \includegraphics[scale=0.27]{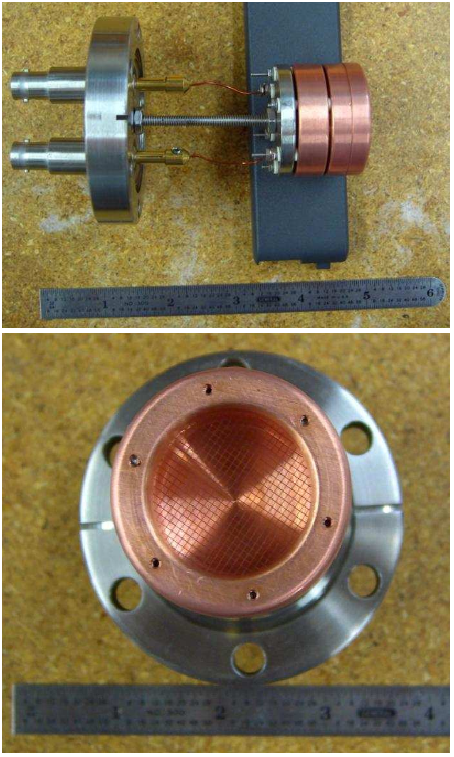}
  \includegraphics[scale=0.23]{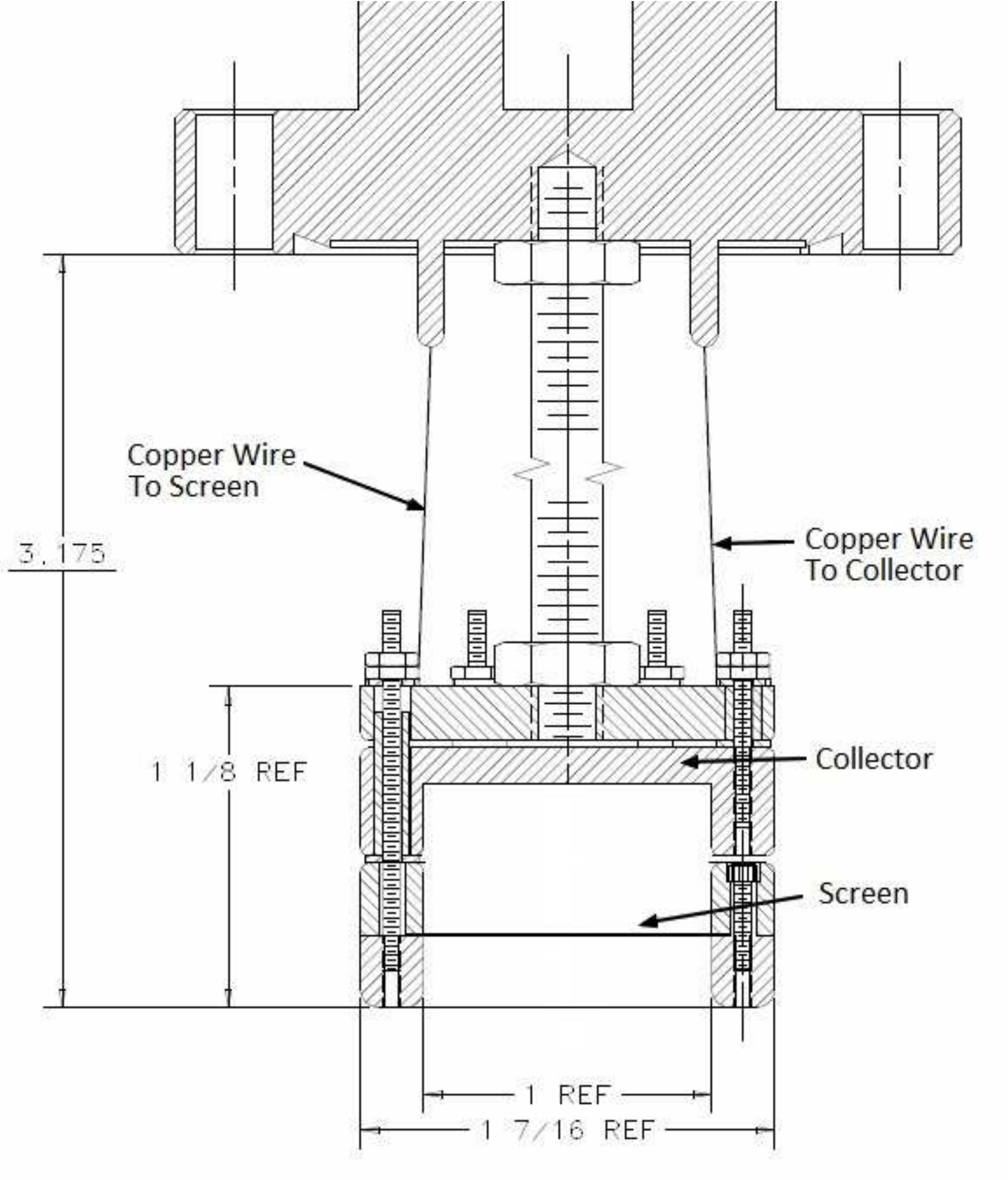}
  \caption{(left) Two views of the RFA used in our setup. A graphite coating is applied to the RFA before installation. (right) A schematic of the RFA used in our setup (electrons enter from the bottom).}
  \label{RFA}
\end{figure}

RFA4 is directly across from RFA1 to serve as a basis of comparison between two RFA designs. RFA4 uses an original Rosenberg design that was used in the MI in 2006~\cite{Xiaolong2007}, whereas the other three RFAs use a design with improved collector sensitivity~\cite{tanRFA2007}. Close-up pictures and a detailed schematic of the improved RFA can be seen in Fig.~\ref{RFA}. 

The beampipe in front of each RFA has slots cut into it so that the RFA is screened from the beam wakes but electrons can pass through the slots to be collected. The collector cup is connected to a 1 M$\Omega$ resistor, a 40 dB preamplifier (preamp) and a low-pass filter with a $3$ dB attenuation point at 3 kHz ~\cite{tanRFA2007}. The RFA collector signal is measured as a negative voltage and our preamp range limits that signal to a maximum magnitude of $-10$ volts. Each preamp also has a different baseline voltage which must be subtracted to obtain the true RFA signal. A deviation from the baselines voltage of $-1$~V is equivalent to an electron flux of $\sim\! 10^{7}$ electrons per cm$^{2}$ per second.

Between the beampipe slots and the collector cup, the RFA features a fine mesh grid which can be set to a voltage between 0 and $-500$~V. The grid allows the RFA to discriminate by electron energy, by ensuring that only electrons with energies exceeding the grid potential reach the collector. L. McCuller~\cite{lee2009} found the efficiency of the RFA to be $90$\% with at least $-20$~V on the grid. With potentials of a lesser magnitude, secondary emissions off the collector decrease the collection efficiency ~\cite{lee2009}. Consequently we monitored the electron cloud during each run using a grid voltage of $-20$~V.

\section{Beampipe Coating Performance \& Conditioning}

To properly characterize the electron cloud mitigation performance of the materials, the electron flux needs to be studied over time. Conditioning is the process where the bombarding electrons change the surface chemistry of the beampipe (see \cite{larci,li2013,vallgrenThesis}). As the beampipe conditions its secondary emission yield becomes lower. The lower secondary emission yield will generate a lower electron cloud signal for the same beam conditions. By tracking changes in the relationship between the Main Injector beam intensity and the electron cloud signals, the rate of conditioning of the coated and uncoated beampipe can be compared.

Fermilab uses the ACNET control system~\cite{acnet} and our work relies on the Lumberjack Datalogger module~\cite{datalog} to automatically trigger, read, timestamp, and record the electron cloud signal at each RFA. For each RFA location, the Datalogger module records the maximum electron cloud signal obtained in each Main Injector ramp cycle~\cite{Backfish}. For a smaller subset of Main Injector cycles, we were able to track the full trace of the electron cloud signal over the ramp cycle. By correlating this data with maxima, we were able to calculate the total charge deposited into the beampipe over the entire ramp of all cycles.

\begin{figure}[!t]
   \centering
     \includegraphics[scale=0.65]{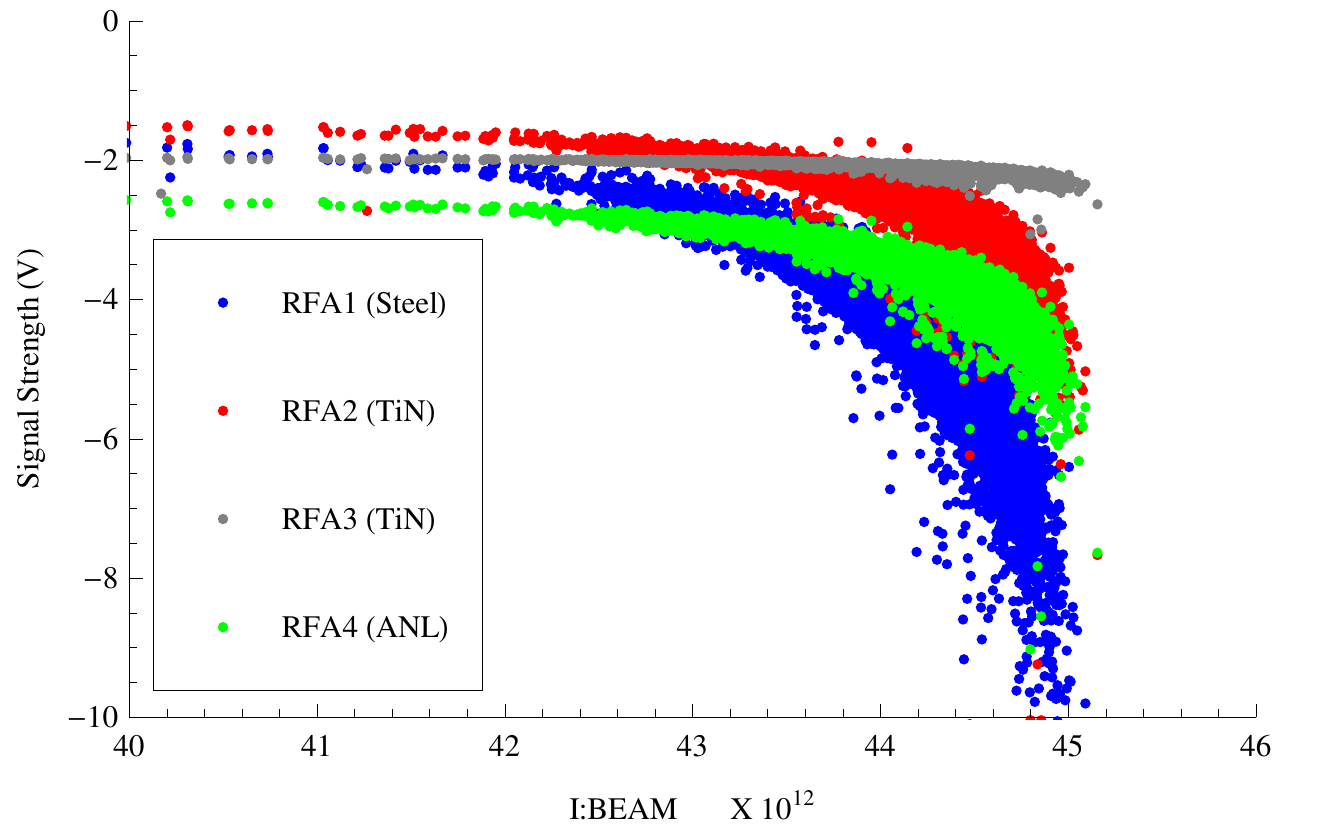}
     \caption{Data collected on March 1, 2010 shown as a scatterplot for each RFA detector. On the horizontal axis is the beam intensity in protons/cycle and on the vertical axis is the RFA signal (maximum per cycle) with negative voltage indicating greater electron flux. The baseline for each signal has not been subtracted. This data was taken after six months of conditioning for the TiN and steel beampipes.}
     \label{TiNcorr}
\end{figure}

In this electron cloud density regime, electron flux has a quasi-exponential dependence on beam intensity. In order to make a meaningful comparison between sets of electron cloud signals, it is essential to correlate the signal recorded at each cycle with the Main Injector beam intensity. A scatterplot showing the electron cloud signal as a function of beam intensity can be seen in Figure~\ref{TiNcorr}.
    
\begin{figure}[!t]
   \centering
     \includegraphics[scale=0.17]{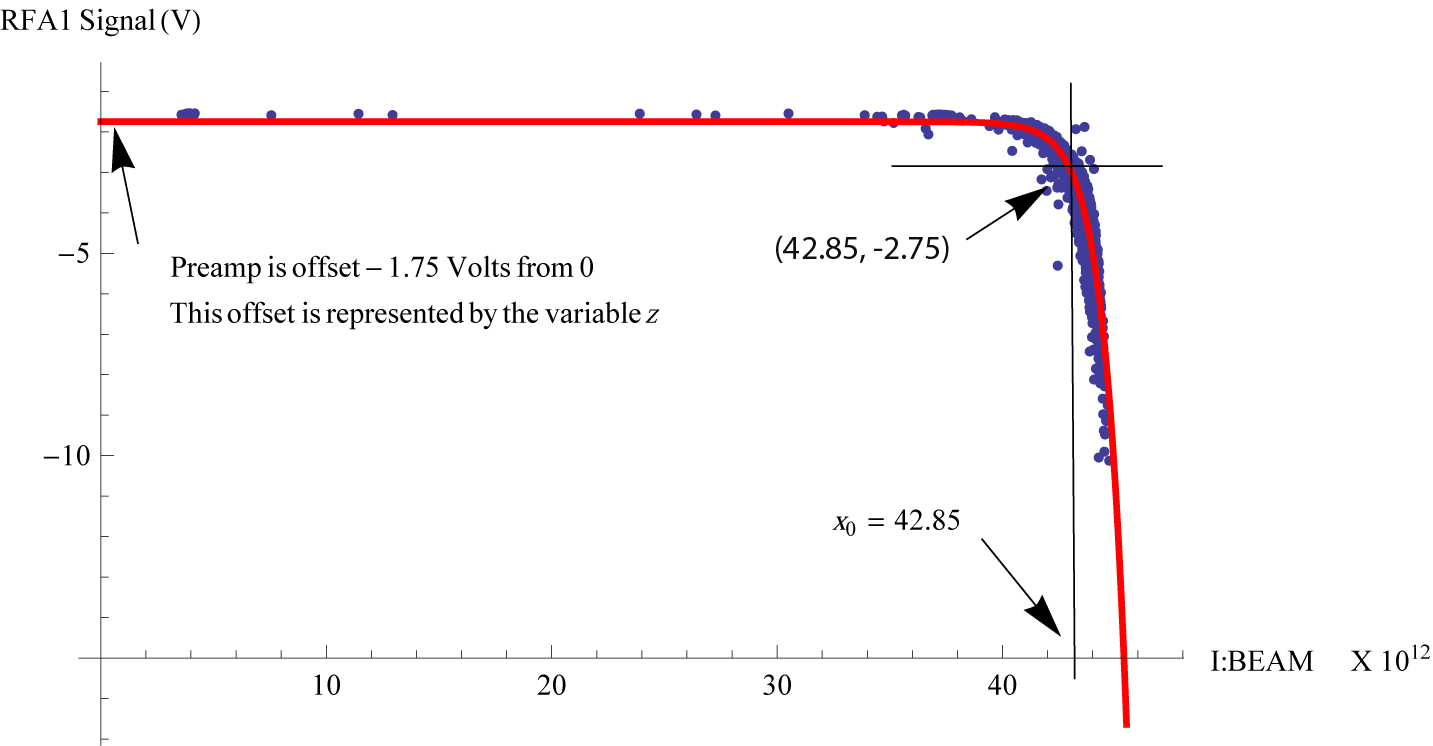}
     \caption{A typical fit of the beam intensity and RFA signal scatterplot is shown. In this case a $-1$~V signal corresponds to a benchmark beam intensity of {$42.85\times 10^{12}$} protons.} 
     \label{Fit}
\end{figure}

We fit the beam intensity and RFA signal scatterplot with an exponential function to characterize the state of the electron cloud within an interval of time. We use the exponential equation in the form
\begin{equation}
V(x)=z_{0} - V_{0} e^{a(x-x_0)}
\label{eqn1}
\end{equation}
where $V(x)$ is the RFA signal is Volts, $x$ is the beam intensity, $z_{0}$ is the baseline voltage, $a$ is the exponential growth parameter, and $-V_{0} e^{-ax_{0}}$ is the exponential coefficient. We define $x_{0}$ to be our beam intensity benchmark and $V_{0}$ to be 1 volt. When the beam intensity $x$ is equal to $x_{0}$, the RFA signal passes a $-1$~V threshold equivalent to a flux of approximately $10^{7}$ electrons per cm$^{2}$ per second. This benchmark serves as a standard means of comparison with which to track the conditioning of the steel beampipe and the coated beampipe over time. Fig.~\ref{Fit} illustrates $V(x)$ fitted to a typical daily dataset.

\subsection{First Run: Titanium Nitride}

TiN has been found to reduce electron cloud in KEKB LER~\cite{kato2010} and has also been shown to inhibit multipactoring~\cite{kuchnir}. The coated beampipe used in our first run was prepared at Brookhaven National Lab via TiN magnetron sputtering~\cite{valerio2009}.

%inhibit multipactoring~\cite{kuchnir}\cite{abe2005}\cite{kaabi2010}.
%multipactoring is electron cloud with a frequency-domain excitation~\cite{costa2012}\cite{montero}.

The beampipe test station at MI-52 was first installed with a TiN-coated section of beampipe alongside an uncoated section of beampipe not previously exposed to beam. On September 11, 2009, the first Main Injector beam operations began following the Fermilab summer shutdown in which the test setup was installed. The beampipe test was continued until July 4, 2010, when the TiN beampipe was removed. In the first month of the run the RFA3 (TiN) signal goes from $\sim$50\% to $\sim$20\% of the RFA1 (steel) signal. By the end of the run, the RFA3 (TiN) signal is 5--10\% of the RFA1 (steel) signal.

\begin{figure}[!t]
   \centering
     \includegraphics[scale=0.4]{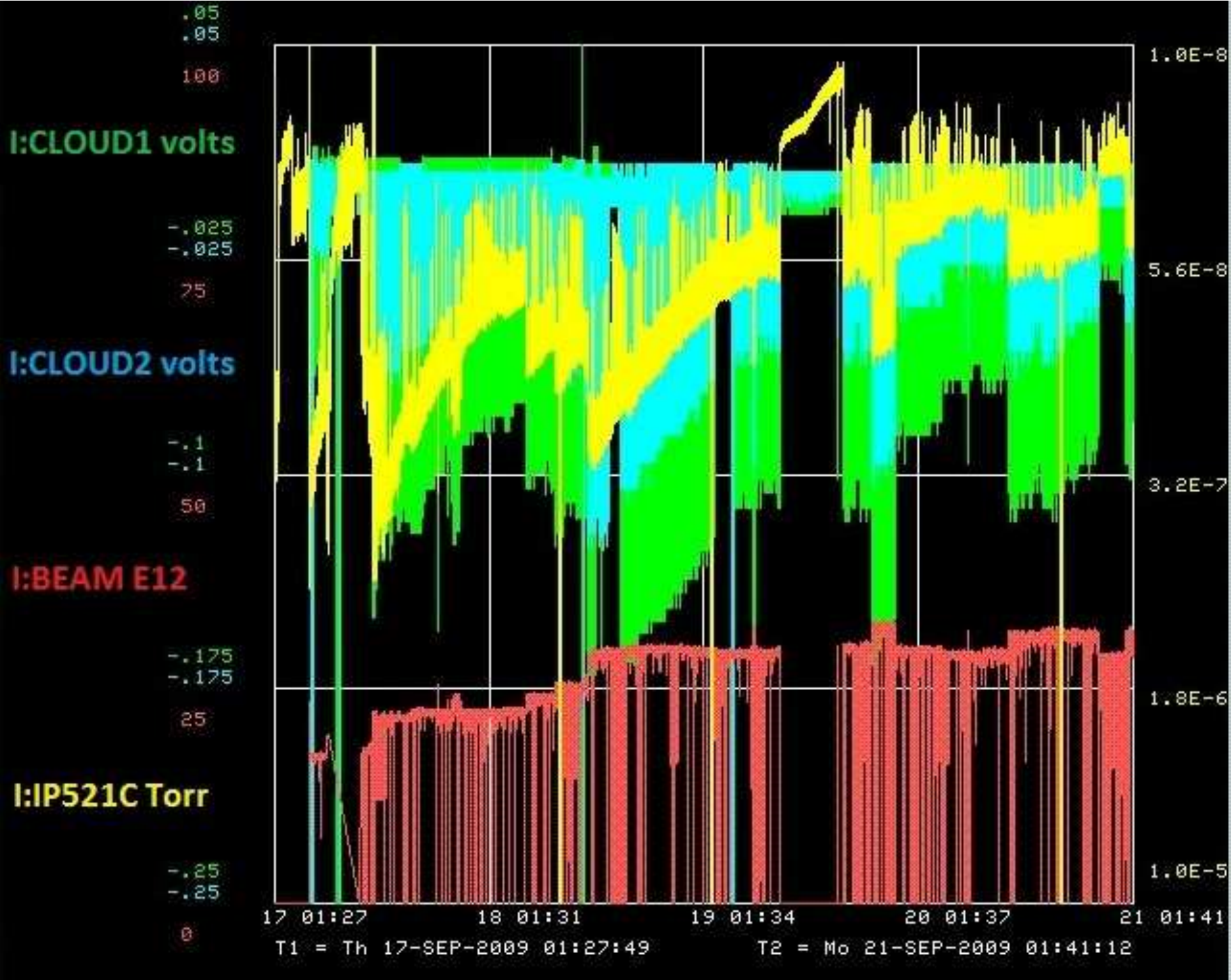}
     \caption{The RFA1 signal (green, I:CLOUD1), the RFA2 signal (blue, I:CLOUD2) and beam intensity (red, I:BEAM) are plotted over time using an Acnet interface. The beampipe conditioning is inferred from the RFA signals decreasing in magnitude as the beam intensity remains constant. The vaccum pressure recorded by a nearby ion pump is plotted on a log-scale (yellow, I:IP521C) alongside the others. The elevated vacuum pressure is caused by the outgassing that occurs as the electron cloud bombards the beampipe.}
     \label{AcnetCond}
\end{figure}

In an early period of data collection, from September 15 to September 25, the slip-stacking signals were too strong to be observed with the preamps turned on. During this interval, the preamp and low-pass filter were bypassed. Fig.~\ref{AcnetCond} shows the RFA signal from Sept. 17th to Sept. 21st. This early period of rapid conditioning correlates with the reduction in vacuum pressure.

\begin{figure}[!t]
   \centering
     \includegraphics[scale=0.4]{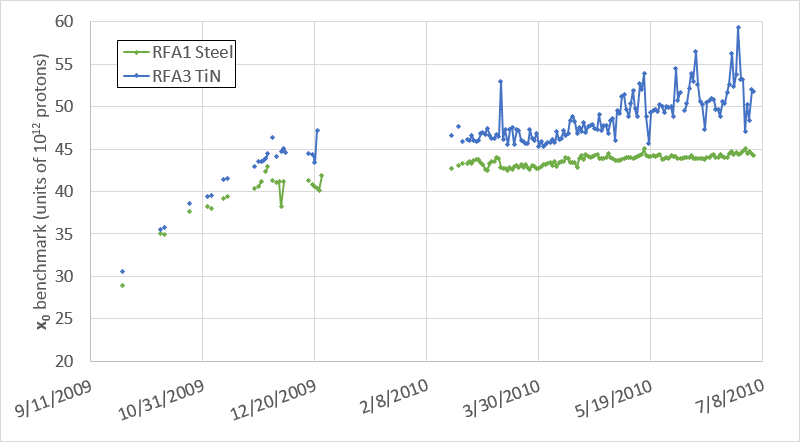}
     \caption{The daily fit of the $x_{0}$ benchmark is shown over time for RFA1 (steel) and RFA3 (TiN). When the $x_{0}$ benchmark passes above 50, the RFA3 signal vanishes because the benchmark exceeds maximum beam intensity currently obtainable in the Main Injector. Therefore, the uncertainty in the $x_{0}$ benchmark for the RFA3 data becomes larger after May 2010. There are days missing benchmark values because $x_{0}$ could not be found from the available data.} 
     \label{TiNthr}
\end{figure}

The beam intensity and RFA signal data was fit each day and the benchmark $x_{0}$ is plotted in Fig.~\ref{TiNthr}. For some days available data could not be fit due to a limited range of our preamps, data acquisition errors, or an insufficient spread in beam intensities. The $x_{0}$ benchmark for these days can be extrapolated based on a local linear regression over days with reliable exponential fits. 

\begin{figure}[!t]
   \centering
     \includegraphics[scale=0.4]{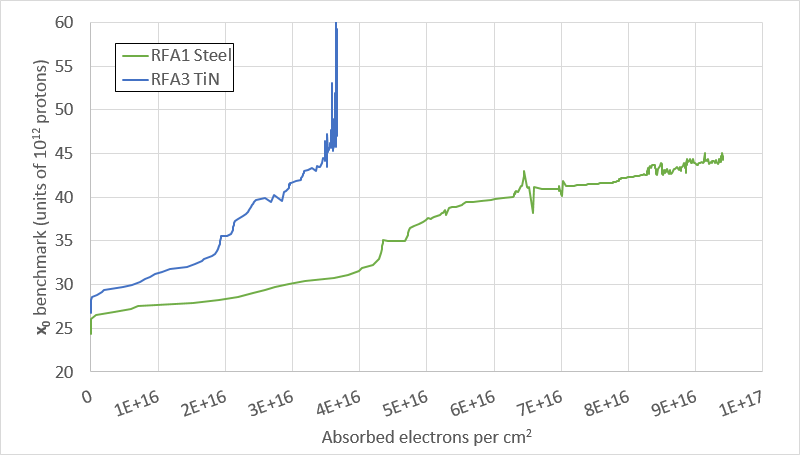}
     \caption{The daily fit of the $x_{0}$ benchmark is shown for RFA1 and RFA3 as a function of the calculated number of electrons absorbed per cm$^{2}$. In the final benchmark values for RFA3, the uncertainty is large.}
     \label{TiNcond}
\end{figure}

Fig.~\ref{TiNcond} shows the benchmark $x_{0}$ as a function of the accumulated number of absorbed electrons. A Lorentzian fitting procedure, described in detail in \cite{Backfish} was used to calculate the integrated RFA signal from the maximum RFA signal in a cycle.

%At any given time, the electron cloud flux at the uncoated beampipe at a given beam intensity is comparable to the flux at TiN coated beampipe at a 10\% higher beam intensity.
The TiN coated beampipe (RFA3) conditioned at a comparable rate per unit time as the uncoated beampipe (RFA1), but did so as a consequence of far fewer number of absorbed electrons. Fig.~\ref{TiNcond} also shows that the steel beampipe (RFA1) is still conditioning after a year of high-intensity running.

\subsection{Second Run: Amorphous Carbon}

The use of {a-C} beampipe coatings for electron cloud mitigation were found to be completely successful at the CERN SPS~\cite{vallgren2011} and this group prepared the {a-C} coated beampipe for our second run. Yin Vallgren's dissertation~\cite{vallgrenThesis} provides a detailed description of preparation of the {a-C} coated beampipe, the theory behind the SEY of carbon materials, and aging studies of {a-C} coated beampipe.

During the 2010 summer shutdown at Fermilab, the TiN coated beampipe was replaced with the {a-C} coated beampipe and the stainless steel beampipe was replaced with unconditioned steel beampipe. The second run of the beampipe coating test station began with beam operations on August 23, 2010 and continued until January 10, 2011. During second run, higher intensity beam occurred sooner than the first run and consequently the beampipes conditioned more rapidly in the first month of the second run.

\begin{figure}[!t]
   \centering
     \includegraphics[scale=0.4]{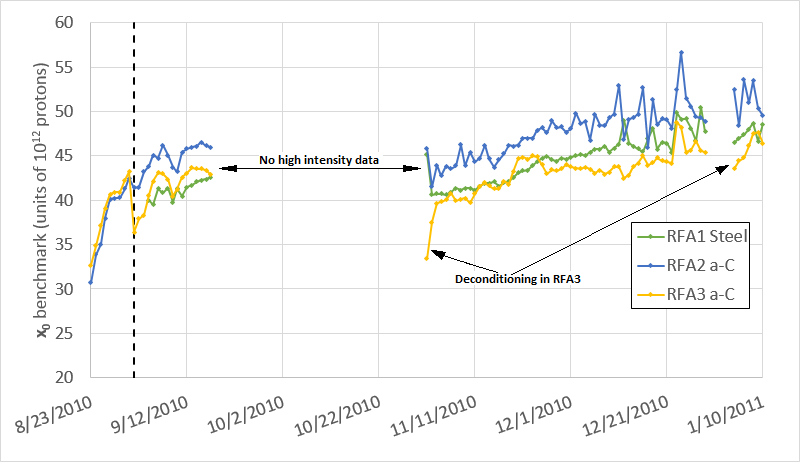}   
     \caption{The daily fit of the $x_{0}$ benchmark is shown over time for RFA1 (steel), RFA2 ({a-C} near steel), and RFA3 ({a-C}). After a vacuum leak occurs (dashed line) near to RFA3, the $x_{0}$ benchmark for RFA3 decreases dramatically and lags behind RFA2.}
     \label{aCthr}
\end{figure}

The benchmark $x_{0}$ from the daily exponential fit is plotted over time in Fig.~\ref{aCthr}. On August 31, 2010, an early vacuum leak near the downstream (RFA3) end of the installation resulted in a dramatic change in the apparent conditioning of the {a-C} beampipe. The benchmark recorded at RFA3 decreases suddenly and lags relative to the benchmark recorded 41 cm upstream at the RFA2 location. Subsequent periods of low intensity are marked by a sharp decrease in the $x_{0}$ benchmark for RFA3. For most of the run, the {a-C} coated beampipe exposed to the vacuum leak (RFA3) performs worse than or comparable to the uncoated beampipe (RFA1), whereas the {a-C} coated beampipe farther from the vacuum leak (RFA2) performs consistently better.

From August 23 to September 4, the RFA1 signal exceeded the range of the preamps and a reliable fit was not obtained. The value of RFA1 at high intensity is extrapolated from its relationship to RFA4 and RFA3 at lower intensities.

\begin{figure}[!t]
   \centering
     \includegraphics[scale=0.4]{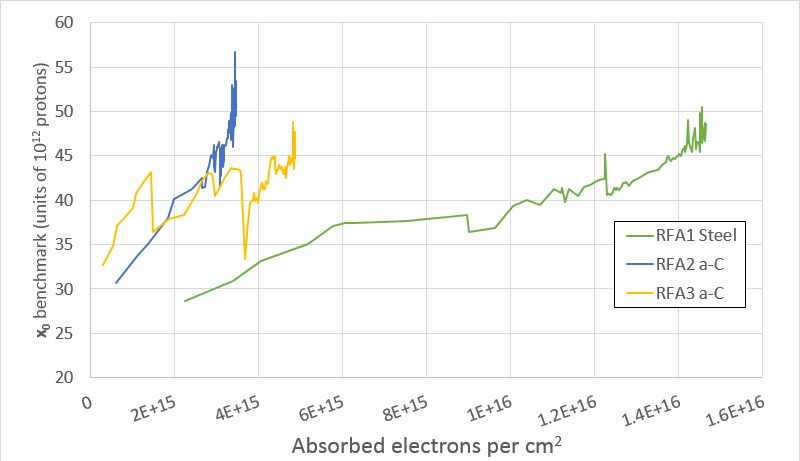}   
     \caption{The daily fit of the $x_{0}$ benchmark is shown for RFA1 and RFA3 as a function of the calculated number of electrons absorbed per cm$^{2}$. In normal operating conditions, the {a-C} coated beampipe near RFA3 conditions at rates comparable to the {a-C} coated beampipe near RFA2. After the vacuum leak or intervals of low beam intensity, the $x_{0}$ benchmark at RFA3 dramatically declines.}
     \label{aCcond}
\end{figure}

Fig.~\ref{aCcond} shows the benchmark $x_{0}$ as a function of the accumulated number of absorbed electrons. The extrapolated values for RFA1 were included in the calculation of the number of absorbed electrons. The Lorentzian fitting procedure \cite{Backfish} was used to calculate the integrated RFA signal from the maximum RFA signal in a cycle.

The {a-C} beampipe was not baked after installation or after the vacuum leak was observed. After the {a-C} beampipe was removed, a visual inspection did not find any discoloration or damage. The contamination currently on the surface of the {a-C} beampipe may be affected by the removal of the beampipe from the tunnel.

\subsection{Third Run: Diamond-like Carbon}

DLC coatings have been tested for electron cloud mitigation at KEKB LER~\cite{kato2010} and CERN SPS~\cite{vallgrenIPAC}. The SEY of DLC is directly compared to other materials in \cite{kato2009}. The KEKB group prepared the DLC coated beampipe for our third run.

During the 2013 summer shutdown at Fermilab, the {a-C} coated beampipe was replaced with the DLC coated beampipe and the stainless steel beampipe was replaced with unconditioned steel beampipe. The third run does not yet contain any beam intensity data above {$28 \times 10^{12}$} protons because the run period coincided with the commissioning of the Fermilab Recycler. After the first two weeks of the third run, no conditioning was observed in any sample. The intensity benchmark $x_{0}$ stayed below {$30 \times 10^{12}$} protons and depended upon beam quality (bunch length, spot size, and vacuum conditions). We will continue this experiment for another year in order to track the conditioning behavior of the DLC at higher beam intensities.

\begin{figure}[!t]
   \centering   
     \includegraphics[scale=0.33]{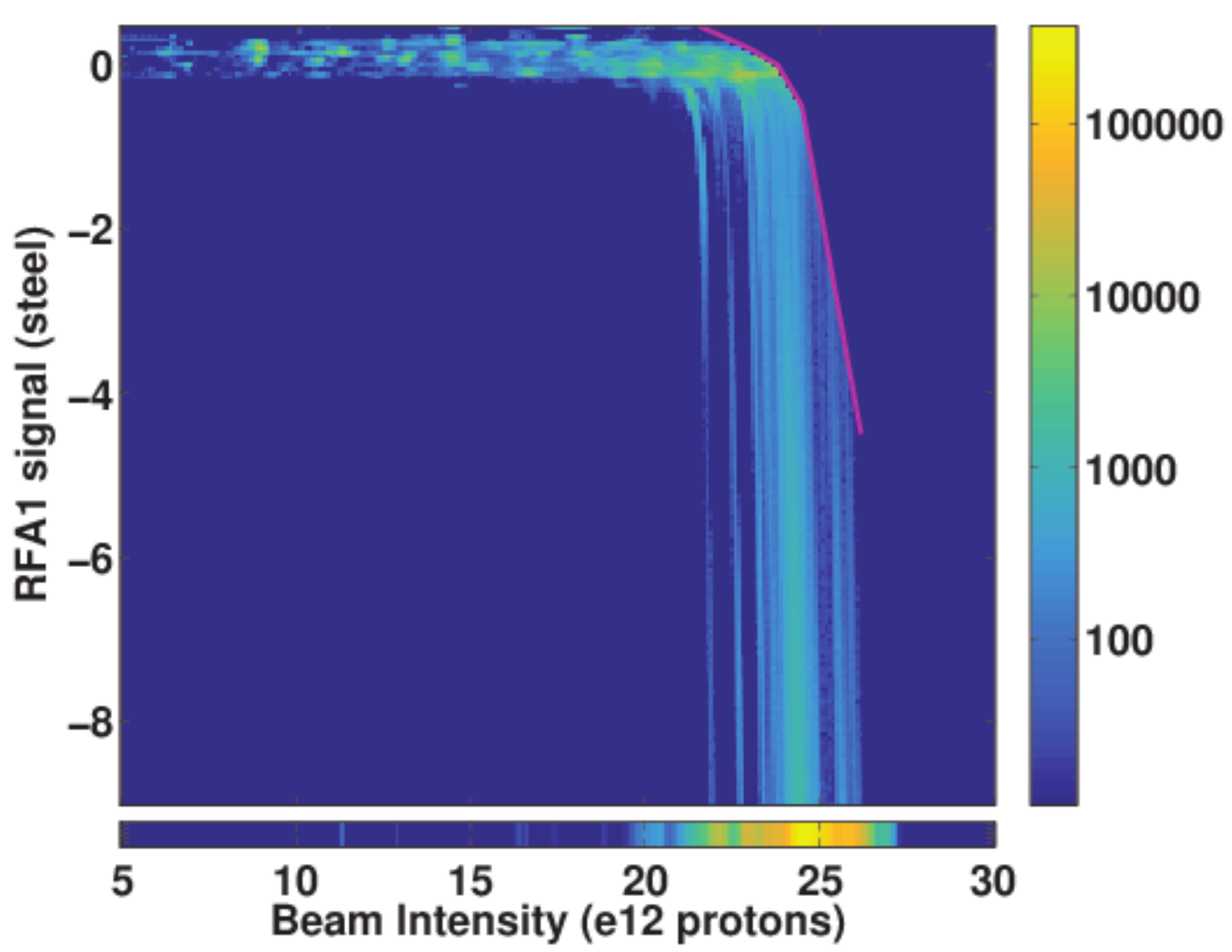}   
     \caption{2D histogram of the RFA1 (steel) signal as a function of beam intensity. Most of the RFA1 signals exceeded the voltage limits of the preamps and this data is shown below the main plot.}
     \label{corrSS}
\end{figure}

\begin{figure}[!t]
   \centering   
     \includegraphics[scale=0.33]{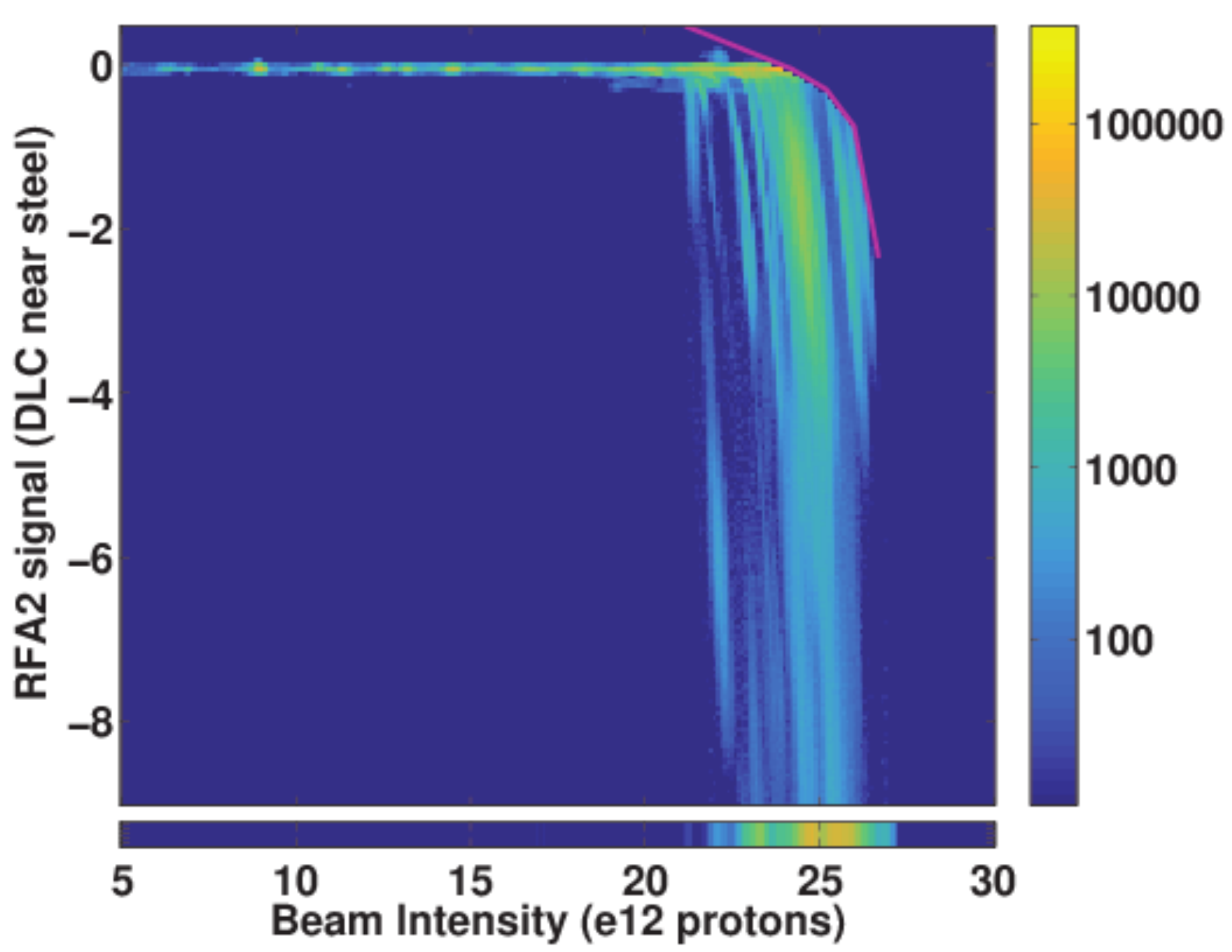}   
     \caption{2D histogram of the RFA2 (DLC near steel) signal as a function of beam intensity. Some of the RFA2 signals exceeded the voltage limits of the preamps and this data is shown below the main plot.}
     \label{corrTRANS}
\end{figure}

\begin{figure}[!t]
   \centering   
     \includegraphics[scale=0.33]{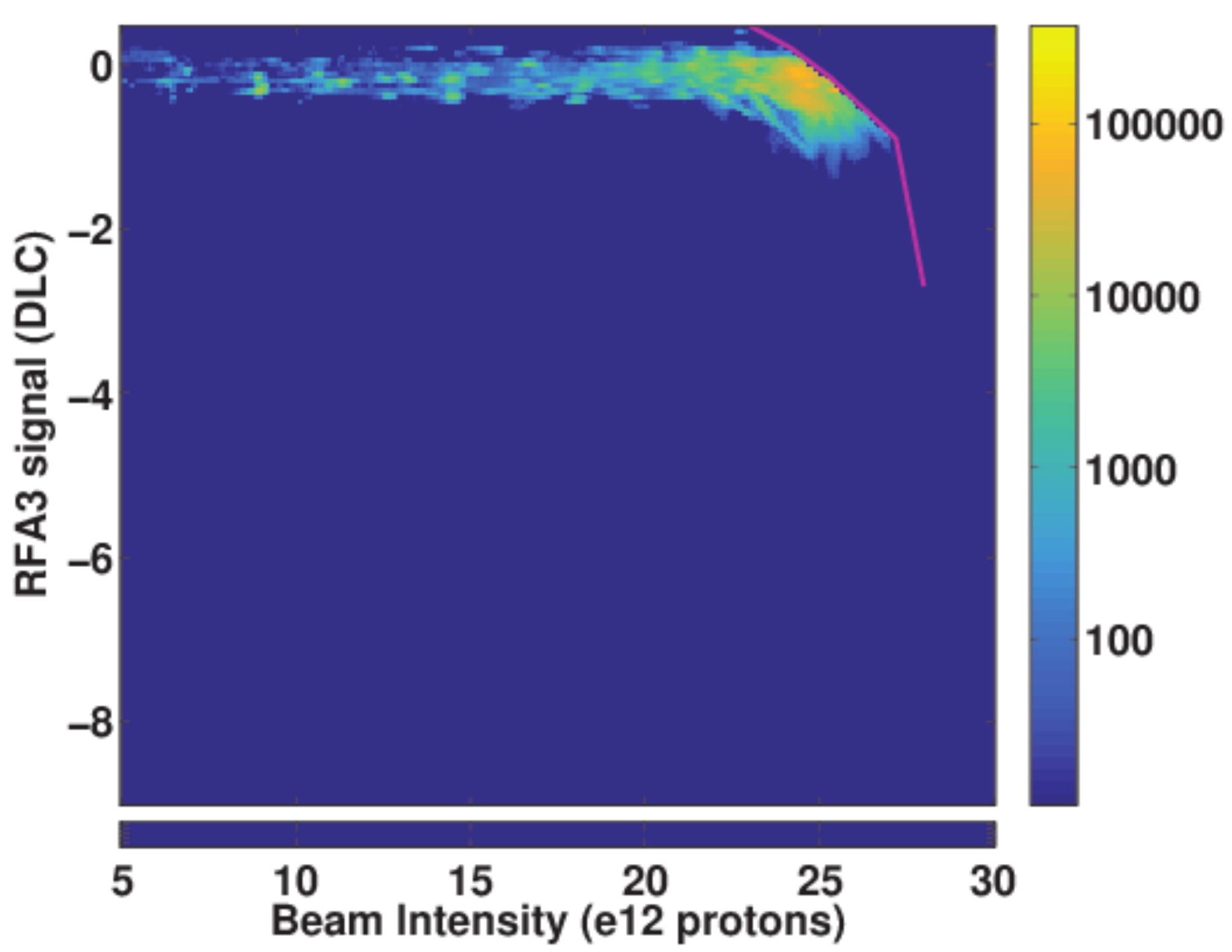}   
     \caption{2D histogram of the RFA3 (DLC) signal as a function of beam intensity.}
     \label{corrDLC}
\end{figure}

Fig.~\ref{corrSS}, Fig.~\ref{corrTRANS}, and Fig.~\ref{corrDLC} show a 2D histogram of the RFA signal as a function of beam intensity for RFA1, RFA2, and RFA3 respectively. The 2D histograms represent the interval from October 1, 2013 to September 10, 2014. There is an artifact from the ACNET Datalogger that occurs when the beam intensity for a cycle with beam is erroneously paired with the RFA signal for a cycle with no beam. These points have been removed from the plots using a piece-wise linear cut (magenta line).

%The data with zero volts and high intensity are an artifact of the ACNET Datalogger that occurs when the beam intensity for a cycle with beam is erroneously paired with the RFA signal for a cycle with no beam.

% The error in the baseline subtraction is comparable to the magnitude of the signal.

For this period, the RFA2 signal is 10--15\% of the RFA1 signal and the RFA3 is 0.5--1\% of the RFA1 signal. The RFA2 signal is primarily seeded by electrons propagated from the denser electron cloud in the stainless steel beampipe 5 cm away. Its not clear whether or not the RFA3 signal is influenced by electrons propagating from the stainless steel beampipe 46 cm away. If the electron cloud attenuates exponentially with length away from the stainless steel, then the RFA3 signal is independent of the signal at RFA1.

%We find the DLC coating to be promising; our future work will examine the conditioning behavior of the DLC coating.

\section{RFA Signal over Ramp Cycle}
\label{MagSec}

\begin{figure}[!t]
   \centering
     \includegraphics[scale=0.4]{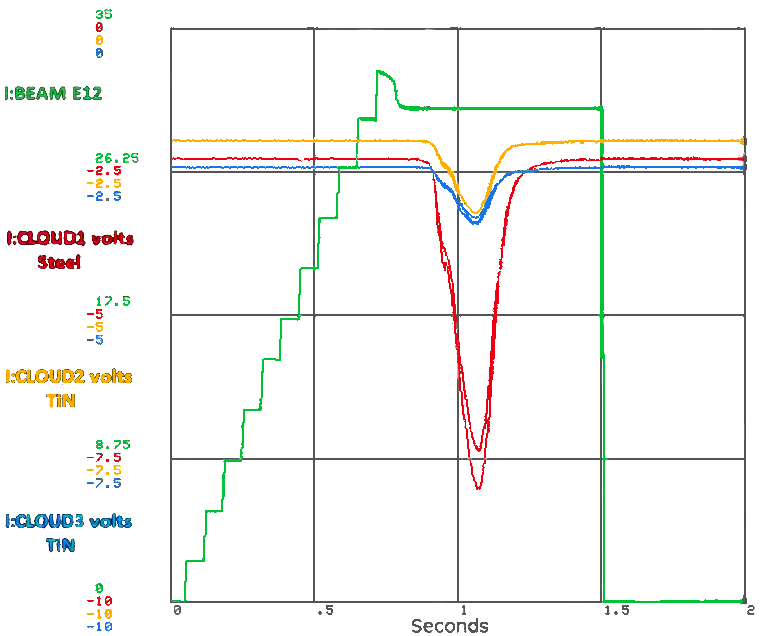}
     \caption{Typical traces of RFA signals over the Main Injector ramp cycle. RFA1 (red, I:CLOUD1), RFA2 (orange, I:CLOUD2), RFA3 (blue, I:CLOUD3), and beam intensity (green, I:BEAM) are shown. The RFA magnitude maximum is obtained near 60 GeV/c.}
     \label{AcnetTiN}
\end{figure}

Fig.~\ref{AcnetTiN} shows the beam intensity and RFA signals for a typical cycle. The RFA signals take a shape empirically similar to a Lorentzian function. The RFA signal maximum is usually obtained at a beam momentum between 50 GeV/c and 70 GeV/c. This strong dependence of the RFA signal on the ramp cycle is not an expected consequence of the beam dynamics. Simulations conducted by Furman~\cite{furman2011} indicate the the change in the RFA signal is not explained by changes in beam spot size or bunch length that occur with the Main Injector ramp.

\begin{figure}[!t]
   \centering
     \includegraphics[scale=0.48]{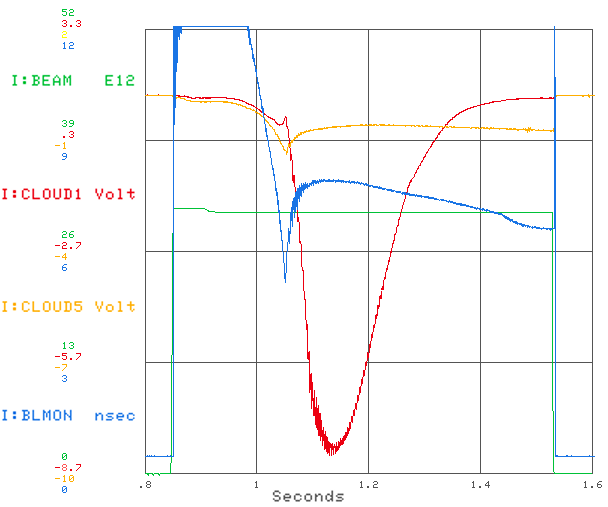}   
     \caption{Beam intensity (green, I:BEAM), RFA1 (red, I:CLOUD1), RFA5 (orange, I:CLOUD2) and Bunch-length (blue, I:BLMON) are shown over a Main Injector ramp cycle from the third run. Transition energy is crossed $\sim$1.05 seconds into the cycle.}
     \label{RFA5}
\end{figure}

Between the second and third run, another RFA, called RFA5, was installed on stainless steel beampipe at MI-10~\cite{rookieMI}. This RFA is part of a separate experiment to make in-situ SEY measurements of small beampipe samples using an electron gun~\cite{scott2012,yichen}. Fig.~\ref{RFA5} indicates the RFA5 signal has a different dependence on the ramp cycle from RFA1. The shape of the signal trace at RFA5 matches what Furman~\cite{furman2011} simulated. The RFA5 signal increases in magnitude as the bunch-length decreases, especially at transition. 

\begin{figure}[!t]
   \centering   
     \includegraphics[scale=0.35]{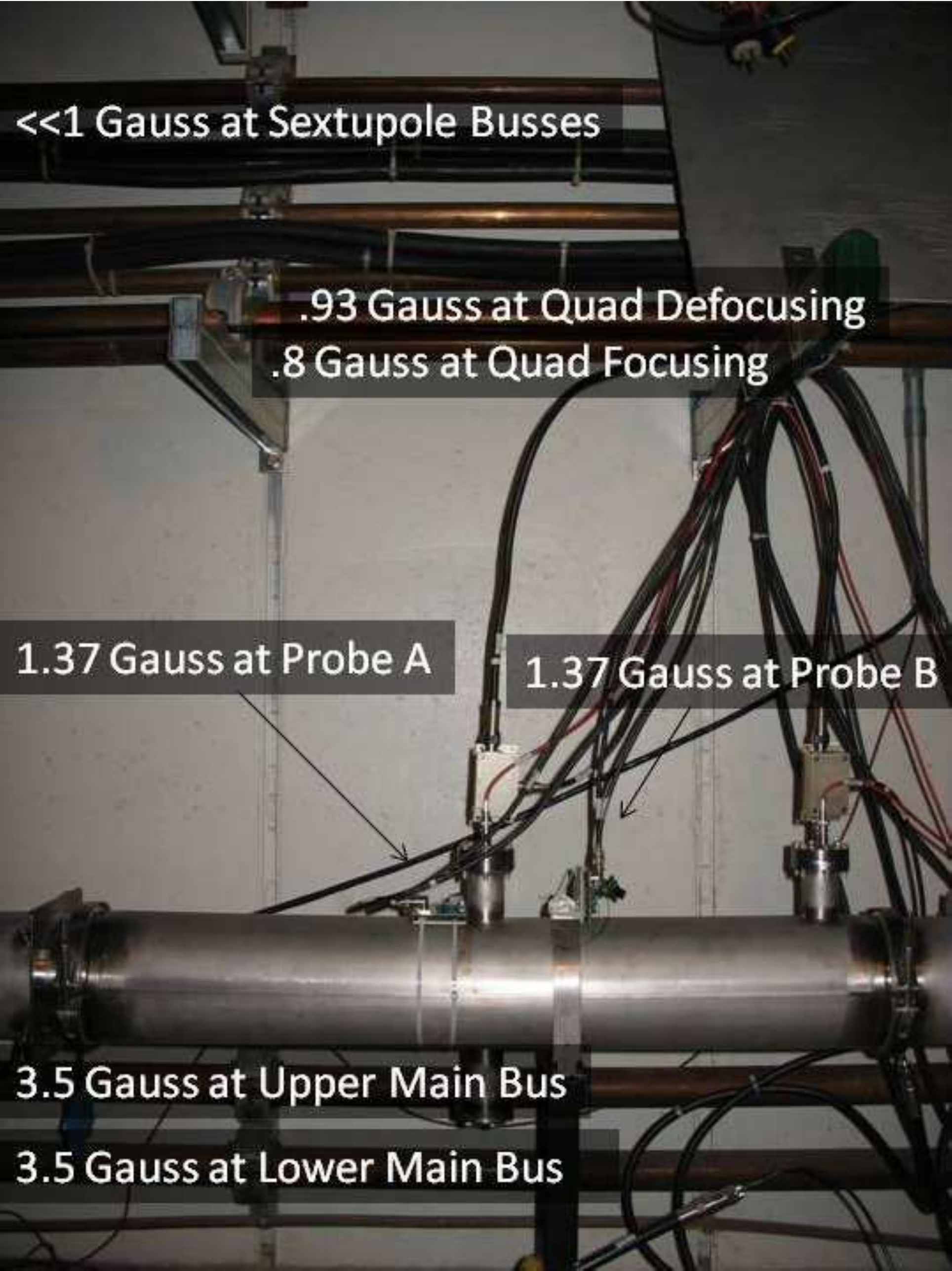}   
     \caption{Residual magnetic fields measured in the Main Injector tunnel at MI-52.}
     \label{BusFields}
\end{figure}

Stray magnetic fields account for the discrepancy between the signal at RFA1 and RFA5. Fig.~\ref{BusFields} shows the nearby magnet buses at MI-52 that give off stray magnetic fields with the Main Injector ramp. Whereas RFA5 is installed at MI-10 near a tunnel alcove where the magnet buses are an order of magnitude farther away. 

\begin{figure}[!t]
   \centering
     \includegraphics[scale=0.35]{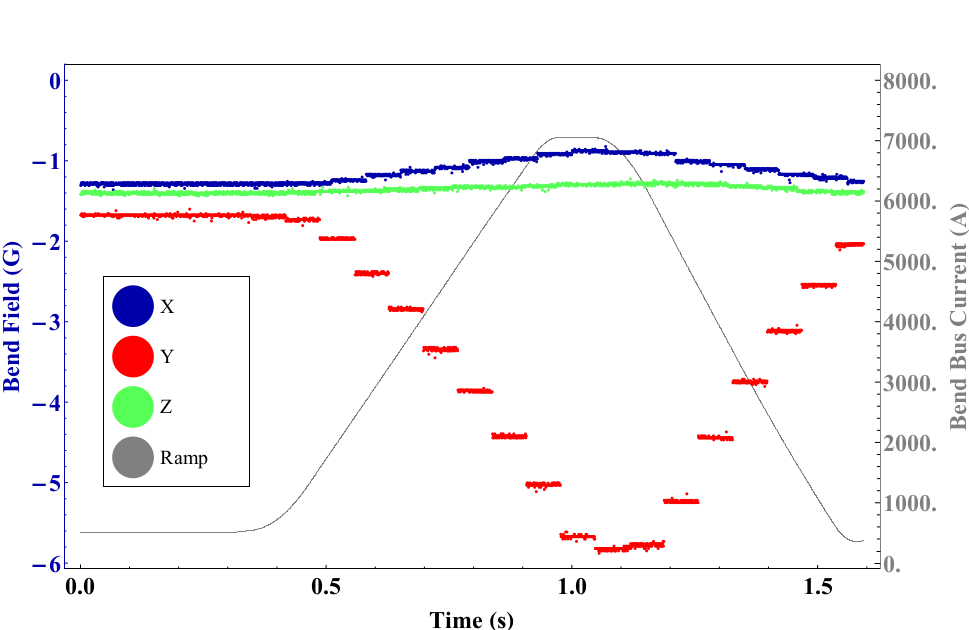}           
     \caption{Stray magnetic fields reach $-5.8$ gauss in the vertical (Y) direction as the bend bus ramps, but are small in the horizontal (X) and longitudinal (Z) directions.}
     \label{Fields}
\end{figure}

Two identical 3-D magnetic Hall probes (``A'' and ``B'') were built to measure the stray fields at MI-52 (3-axis measurements made with \cite{ametes} and magnitude measurements made with \cite{pni}). The field measurements were recorded by the Acnet Datalogger module. Fig.~\ref{Fields} shows the magnetic fields at MI-52 as a function of the Main Injector ramp. Table~\ref{MagTab} compares the peak magnetic fields as measured by each Hall probe, above and below the beampipe. The differences between two probes are not statistically significant but there is a statistically significant difference between the top and the bottom of the beampipe (p=0.001). This indicates the field in the beampipe is not a purely dipole field.

\begin{table}[!t]
\caption{Measurements of the stray magnetic field magnitudes above and below the beampipe using Hall probes A and B.}
\label{MagTab}
\centering
\begin{tabular}{| l | c | c | c|}
\hline
~ & Probe A & Probe B & Combined \\
\hline
Top of Beampipe & $4.9\pm 0.3$ G & $5.6\pm 0.2$ G & $5.25\pm 0.2$ G \\ 
Bottom of Beampipe & $6.4\pm 0.3$ G & $5.9\pm 0.2$ G & $6.15\pm 0.2$ G \\
\hline
\end{tabular}
\end{table}

Simulations indicate that stray fields of less than 10 gauss do not impact the detection efficiency of the RFA~\cite{lee2009}. However fields less than 10 gauss do have a significant impact on the structure of the electron cloud and the fraction of the electron flux oriented vertically towards the RFAs ~\cite{lebrun2013,lebrun2010}. Those simulations find an increase in the RFA signal with the magnetic field, but do not find the decline in the RFA signal at higher magnetic fields. The same electron simulation program (POSINST) was used to analyze the RFA design at CESR~\cite{CalveyRFA}.

RFA1, RFA2, and RFA3 share the same orientation on the beampipe, sources of magnetic fields, and beam qualities. Therefore, the shared features measured by the three RFAs are not a source of uncertainty in the relative performance of the beampipe coatings.

%Therefore any features of the RFA signal that are unaccounted are shared between three RFA locations and not a source of uncertainty in the relative performance of the beampipe coatings.

\begin{figure}[!t]
   \centering
     \includegraphics[scale=0.29]{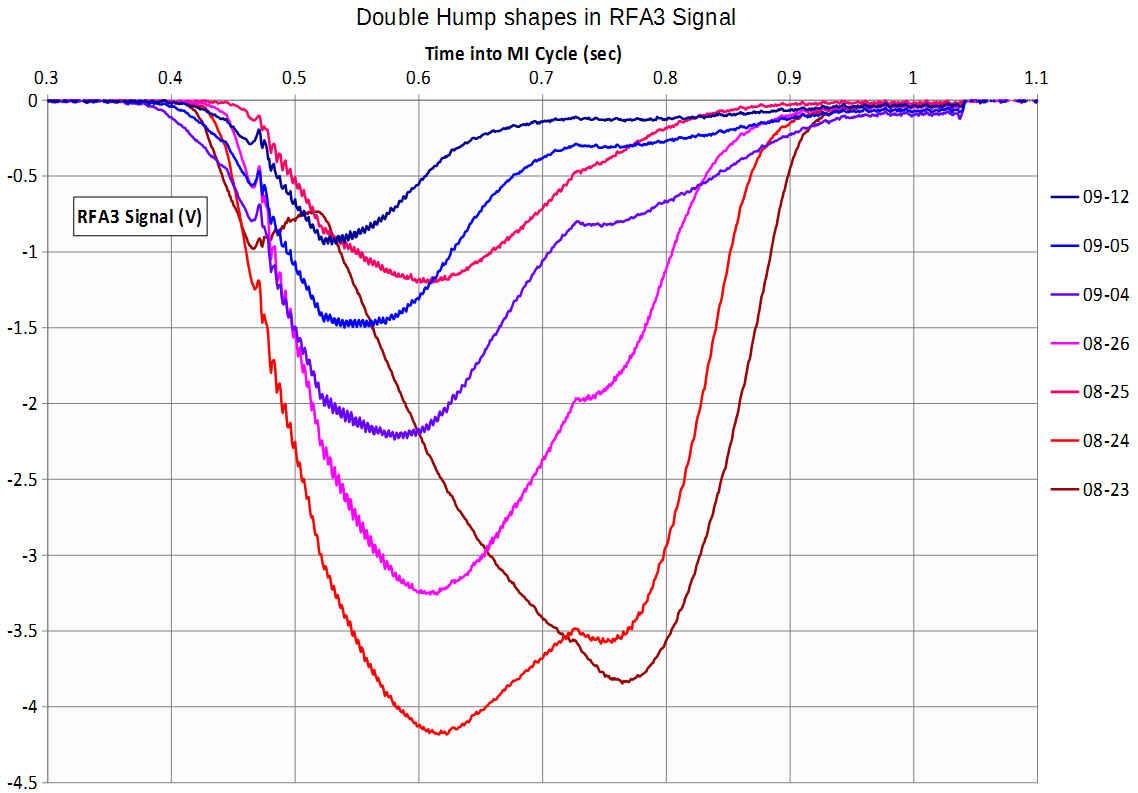}   
     \caption{RFA3 signal over the selected ramp cycles. As the {a-C} conditions, the RFA3 signal becomes weaker and more unimodal. Transition energy is crossed $\sim$0.48 seconds into the cycle.}
     \label{DHshape}
\end{figure}

In the first few weeks of the {a-C} run, RFA3 showed an unusual double-hump shape that is not seen at other RFA locations or in other runs. Fig.~\ref{DHshape} shows selected RFA3 cycles between August 23, 2010 and September 12, 2010. As the {a-C} beampipe conditions, the second maximum gradually disappears into the first one. The origin of this double-hump signal shape, like the single-hump signal shape, is an open question.
	
%Crisp \textit{et al.}~\cite{eddy2009} made measurements of the electron cloud in high-field regions of the Main Injector using a microwave measurement technique. Eldred \textit{et al.}~\cite{eldredHB} made RFA measurements in the Fermilab Recycler, where the Fermilab proton beam is at 8 GeV.

\section{Energy Spectrum of Electron Cloud}

The energy distribution of the electron flux can be inferred by comparing the RFA signal obtained at different grid voltages. The RFA signals are collected over a period of approximately three hours while the grid voltage scans from $-20$~V to $-400$~V in 20~V increments. Since the electron cloud is very sensitive to intensity, we record the maximum RFA signal in each cycle and the corresponding beam intensity for that cycle. For each grid voltage an independent quadratic fit is applied to the beam intensity and RFA signal scatterplot. Next we picked a single beam intensity that is well-sampled at all grid voltages and evaluate the quadratic fit at that point. The difference between the magnitude of this signal at two grid voltages is proportional to the number of electrons with energy between these two energies. For example, the difference between the signal at $-20$~V and $-40$~V represents the number of electrons with energies between 20~eV and 40~eV. We calculate the difference in signal at every 20~V increment and normalize the result to obtain the energy distribution of the electron flux.

\begin{figure}[!t]
   \centering
     \includegraphics[scale=0.45]{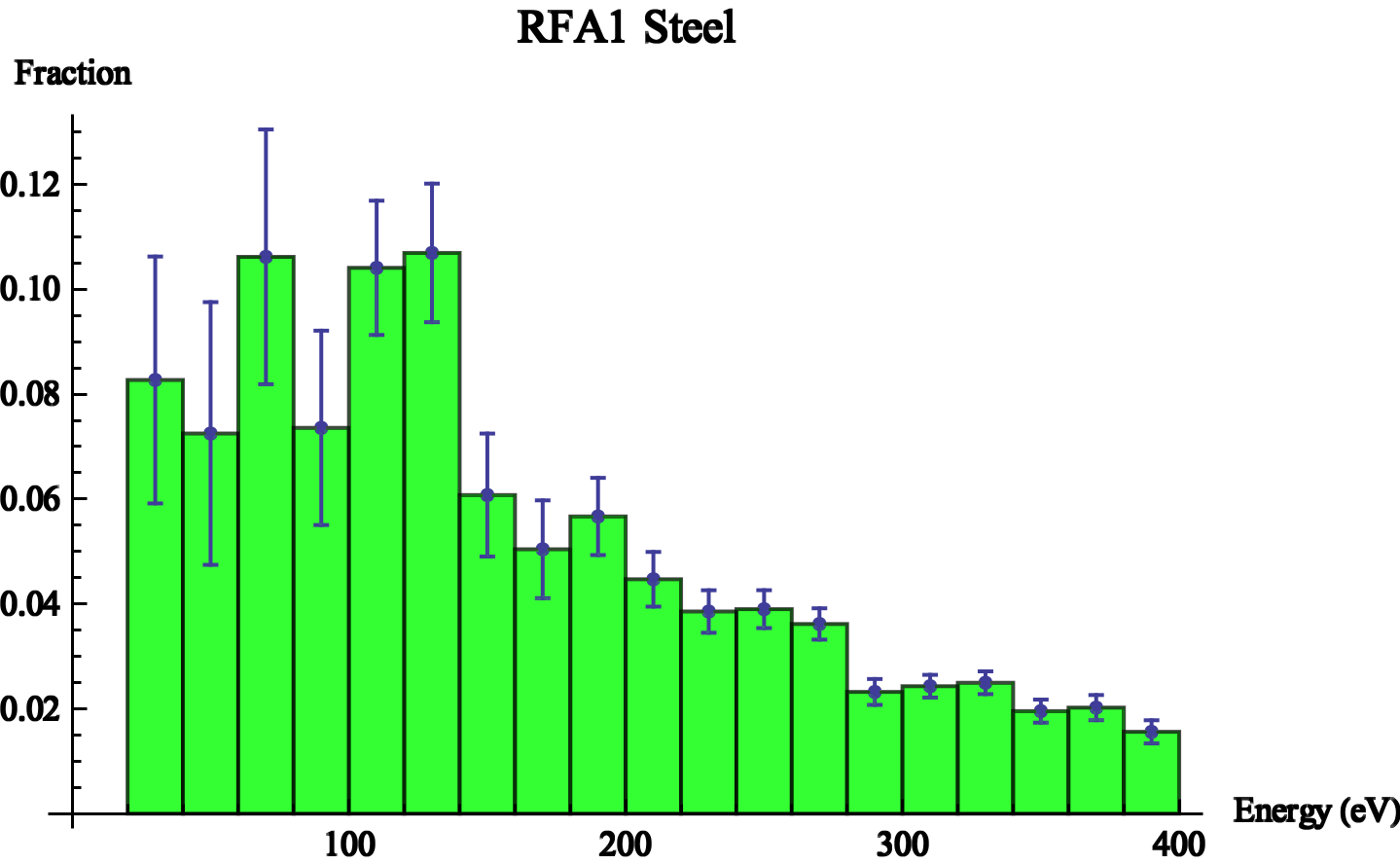}   
     \includegraphics[scale=0.45]{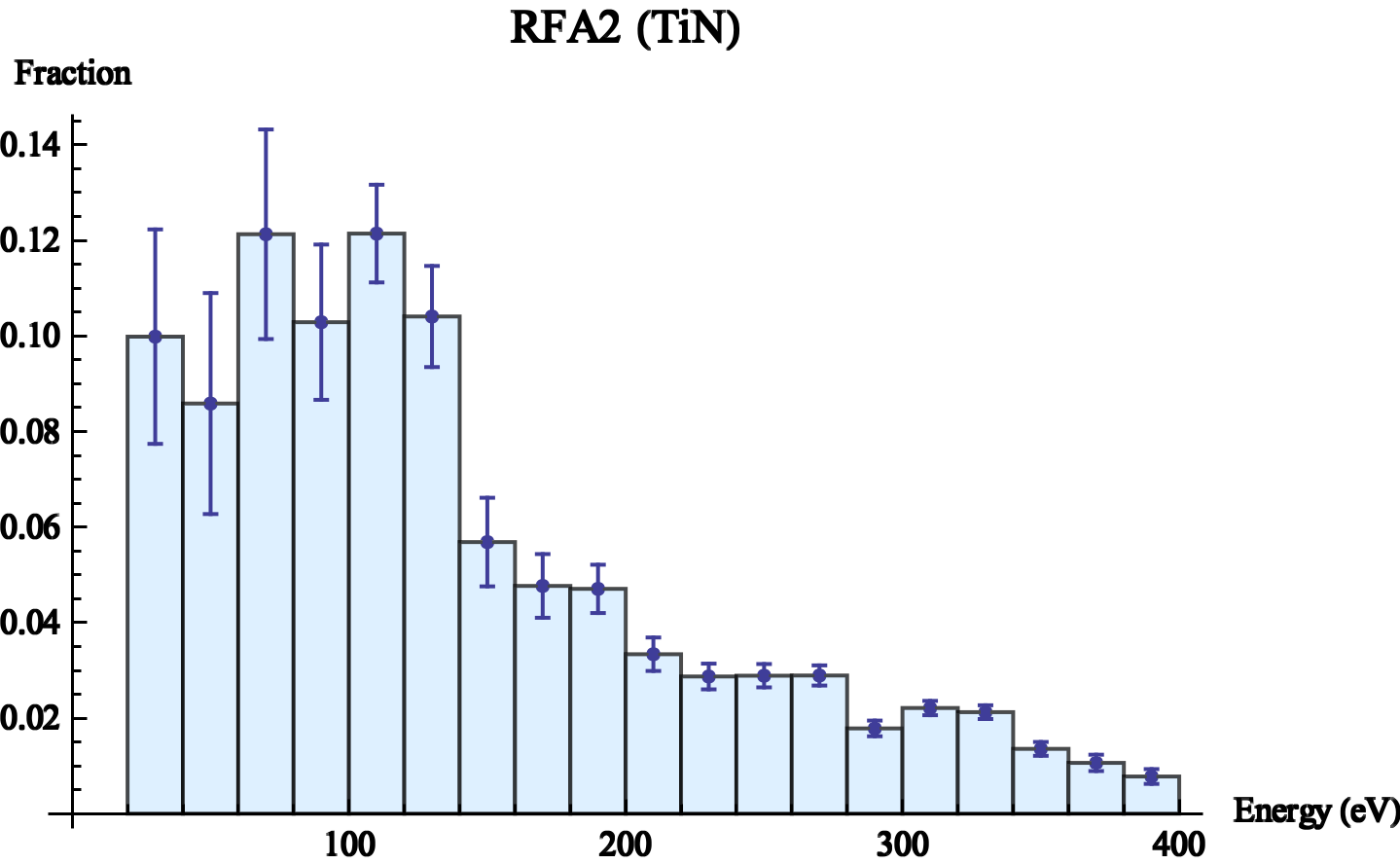}    
     \includegraphics[scale=0.45]{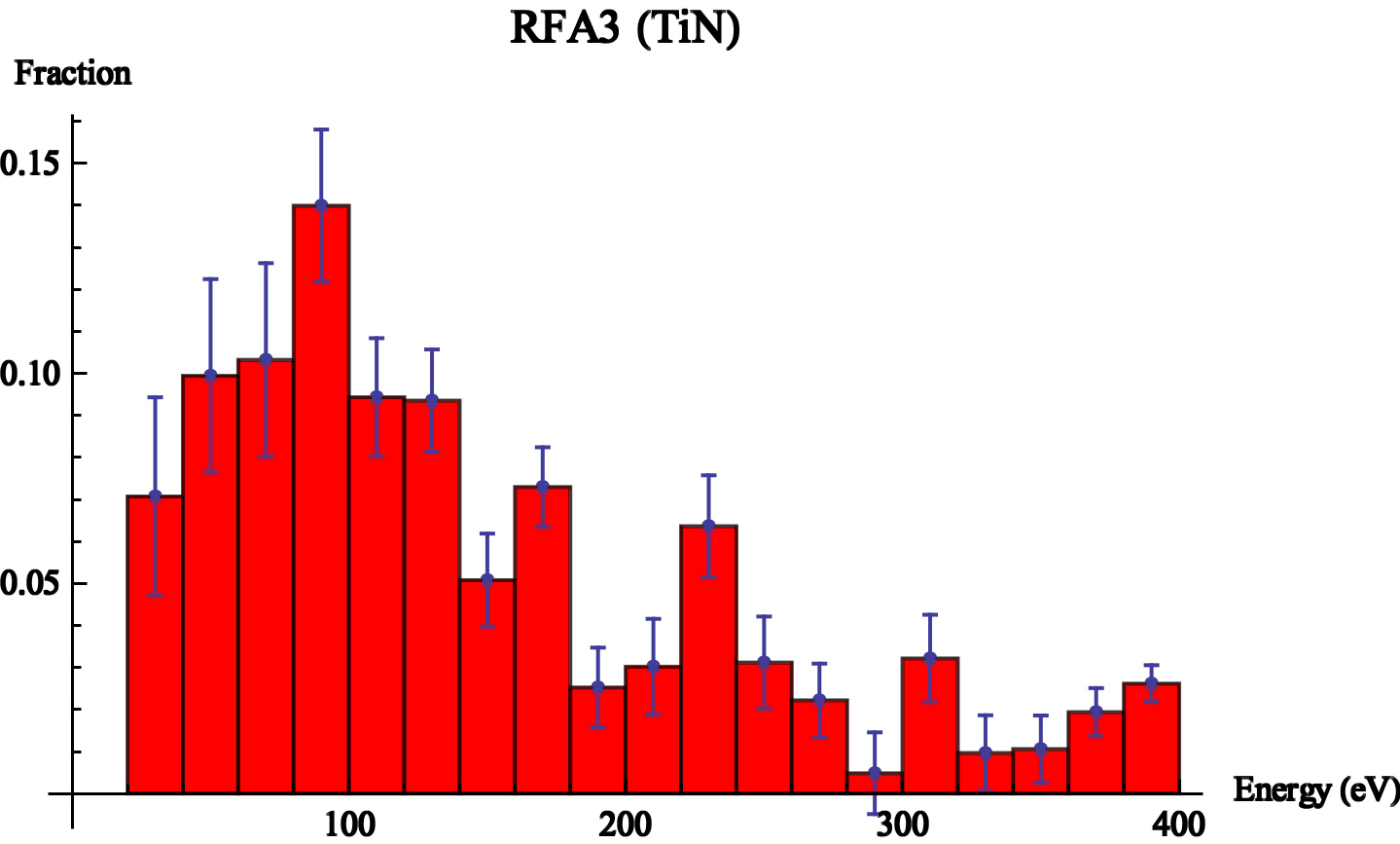}    
     \includegraphics[scale=0.45]{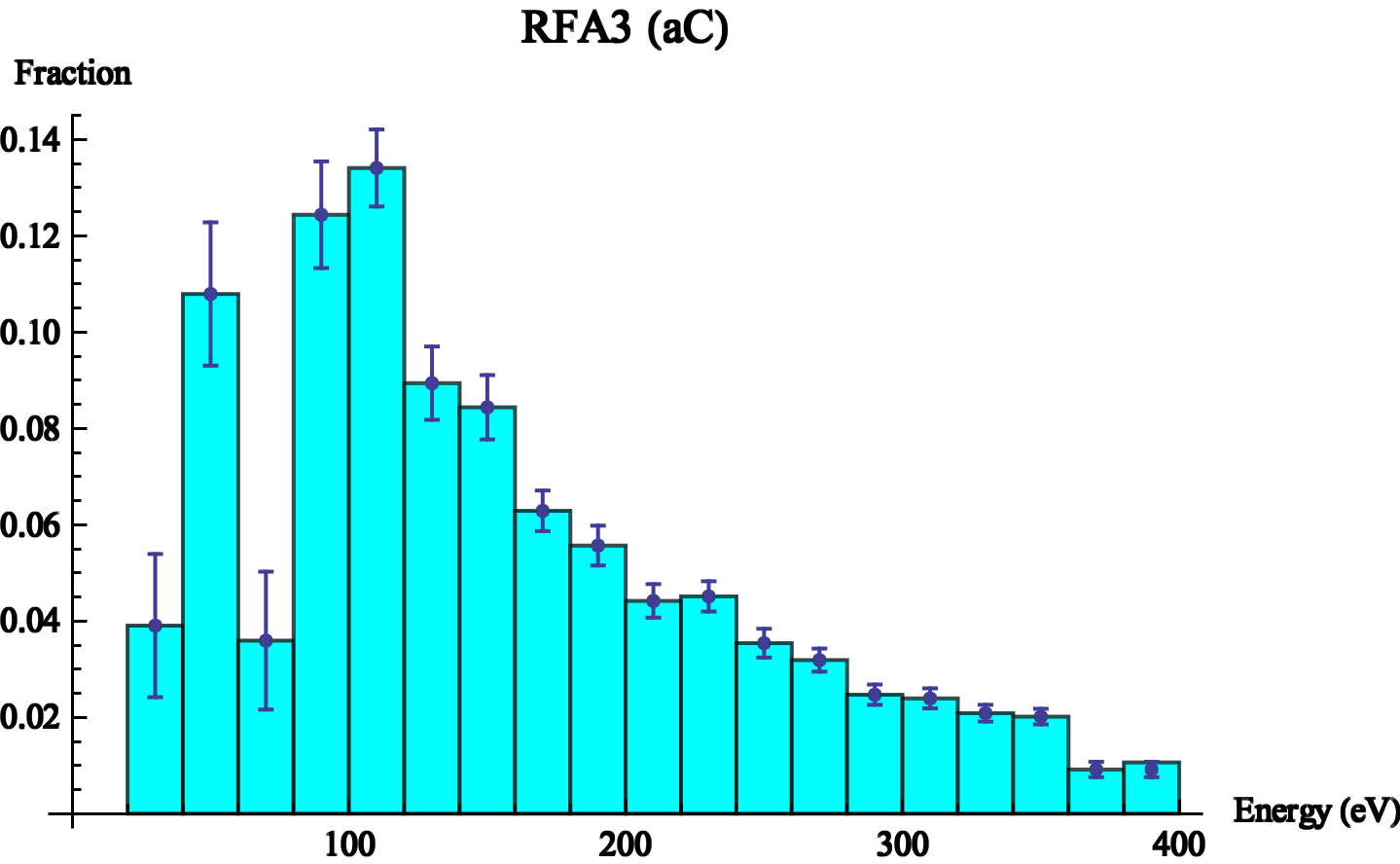}           
     \caption{Electron energy distribution measurement for steel, TiN and {a-C}. The steel and TiN measurements were evaluated a beam intensity of {$42.5\times 10^{12}$} protons per cycle, whereas the {a-C} measurement was evaluated at {$40.0 \times 10^{12}$}.}
     \label{ESpec}
\end{figure}

Fig.~\ref{ESpec} shows the energy distribution of each RFA location during the first run and RFA3 during the second run. Each energy distribution has a peak near 100~eV.

\section{Conclusion}
This work provides an important counterpoint to previous accelerator tests of beampipe coatings for electron cloud mitigation. A fortuitous vacuum leak at the test location suggests that the robustness of the {a-C} coating could be a major concern. Our results indicate that the TiN coating and the (uncontaminated) {a-C} coating have comparable performance at mitigating electron cloud and that both condition more rapidly than the stainless steel. Relative to uncoated steel, these two coatings offer a factor of 6--10 decrease in electron cloud flux at a given beam intensity. Our preliminary results on the DLC coating are very promising -- we find a factor of 100-200 decrease in electron cloud flux at a given beam intensity.
%a 10\% increase in beam intensity for the same electron cloud density or

We provide measurements of the energy distribution of the electron cloud and find strong similarities in distribution generated in steel, TiN, and {a-C} beampipe. We also examine the RFA signal over the course of the ramp cycle and find a strong sensitivity to bunch-length and weak magnetic fields.
%The decline in the electron cloud signal at the end of the ramp cycle has not been anticipated or replicated in simulation.

\section*{Acknowledgment}
The authors would like to thank D.~Capista, L.~Valerio, and the electron cloud team at BNL for preparing the TiN beampipe, C.~Yin~Vallgren and the electron cloud team at CERN for preparing the a-C beampipe, and S.~Kato and the electron cloud team at KEK for preparing the DLC beampipe. The authors would also like to remember the late Tom Droege, whose diligence is attested to in his enduring high-voltage power supplies.

\bibliographystyle{IEEEtran}
\bibliography{bibliography}


\end{document}